\documentclass{INTERSPEECH2023}

\interspeechcameraready

\usepackage{enumitem}
\usepackage{multirow}
\usepackage{pifont}
\newcommand{\cmark}{\ding{51}}%
\newcommand{\xmark}{\ding{55}}%

\title{Multi-Head State Space Model for Speech Recognition}
\name{Yassir Fathullah$^{1,2}$\sthanks{\hspace{1.5mm}Work done during internship at Meta AI.}, Chunyang Wu$^2$, Yuan Shangguan$^2$, Junteng Jia$^2$, Wenhan Xiong$^2$, Jay~Mahadeokar$^2$, Chunxi Liu$^2$, Yangyang Shi$^2$, Ozlem Kalinli$^2$, Mike Seltzer$^2$, Mark J. F. Gales$^1$
}

\address{
  $^1$University of Cambridge, UK \\
  $^2$Meta AI, USA
}
\email{yf286@cam.ac.uk, chunyang@meta.com}

\begin{document}

\maketitle
 
\begin{abstract}
    State space models (SSMs) have recently shown promising results on small-scale sequence and language modelling tasks, rivalling and outperforming many attention-based approaches. In this paper, we propose a multi-head state space (MH-SSM) architecture equipped with special gating mechanisms, where parallel heads are taught to learn local and global temporal dynamics on sequence data. As a drop-in replacement for multi-head attention in transformer encoders, this new model significantly outperforms the transformer transducer on the LibriSpeech speech recognition corpus. Furthermore, we augment the transformer block with MH-SSMs layers, referred to as the Stateformer, achieving state-of-the-art performance on the LibriSpeech task, with word error rates of 1.76\%/4.37\% on the development and 1.91\%/4.36\% on the test sets without using an external language model.
    %
\end{abstract}

\section{Introduction}
\label{sec:introduction}

%
%

Recurrent neural networks (RNNs) have historically been a core approach for a wide range of sequence modelling tasks such as speech recognition \cite{las, rnn-speech}, machine translation \cite{cho-etal-2014-learning, attention} and language modelling \cite{rnn-lm1, rnn-lm2}.
%
However, RNNs were rapidly replaced with the introduction of the transformer \cite{transformer} and large pre-trained models \cite{devlin2018bert}. The effectiveness of the transformer has also caused other fields such as computer vision to consolidate towards attention-based models \cite{vit1, vit2}.

%
One of the key reasons behind the success of transformers and their widespread use is the self-attention mechanism. Unlike previous approaches, self-attention was shown to be highly effective at capturing global features of a sequence by modelling all pairwise interactions.
%
Furthermore, while transformers are exceptionally good at capturing global long-range dependencies they are less able in modelling local patterns. To this end, there has been a range of work on combining convolutional networks with attention for sequence modelling, and has been found to be highly effective for speech recognition \cite{conformer, peng2022branchformer, ebranchformer}.
%

Meanwhile, the deep learning community has slowly been paying more attention to alternative recurrent neural network approaches to effectively and efficiently model sequences \cite{lssl, lmu, lei2018sru, lei2021srupp}. 
Specifically, a well-established signal processing and control theory technique, the state space model (SSM) \cite{kalmanssm}, which historically has been widely used in many continuous or discrete time-series and control problems \cite{ssm-timeseries-book2, ssm-timeseries-book} has been under renewed scrutiny.
However, the recurrent time-variant nature of general SSMs has traditionally made them computationally expensive and inaccessible to many large-scale sequence tasks.
%
Nonetheless, recent work has shown that it is possible to simplify and scale state space models and train them in parallelized manner, by equivalently rephrasing them as a convolution with variable-length kernels \cite{lssl, lmu}. 
Further work has also shown that simplified, structured versions of the time-invariant state space model can be highly efficient, handle long-range dependencies and be a formidable alternative to self-attention on some sequence tasks, such as language modelling \cite{hippo, s4, gss}.

In this work, we evaluate and extend the work on SSMs for speech recognition, proposing a multi-head state space model, equipped with a novel gating mechanism. Since SSMs have shown promising performance on long sequence tasks we hypothesize that a multi-headed approach could better handle both short and long-term modelling simultaneously.
We investigate the use of such multi-head state space models both as a replacement and complement to self-attention in the acoustic encoder of a neural network transducer model.
Our technical contributions include:
\begin{enumerate}[label=(\alph*), leftmargin=5.3mm]
    \item \textit{Stacked and multi-head generalization}. We extend the SSM approach by allowing multi-head processing of linearly projected lower-dimensional signals and stack such a layer for better performance.
    \item \textit{Head gating}. We propose an inter-head gating approach in which different SSMs within the multi-head layer communicate by gating each other.
    \item \textit{Combination with attention}. We also augment the transformer encoder by including a bidirectional SSM residual block prior to the attention block for state-of-the-art performance. This model is referred to as the \textit{Stateformer}.
\end{enumerate}
With these contributions, we advance the state of attention-free models on the LibriSpeech speech recognition task, outperforming strong attention-based baselines. We also show that the Stateformer can achieve state-of-the-art performance on this task with word error rates of 1.76\%/4.37\% on the development and 1.91\%/4.36\% on the test sets, without using an external language model.

\section{State Space Model: The Linear RNN}
\label{sec:background}

The time-invariant state space model \cite{linearssm} is a fully linear recurrent network taking the following form:
\begin{equation}
    \label{eq:ssm}
    \left.\begin{array}{c}
        \begin{alignedat}{3}
            \bm x_k &= {\bm A} \bm x_{k-1} + {\bm B} \bm u_k \\
	    \bm y_k &= {\bm C} \bm x_k + {D} \bm u_k
        \end{alignedat}
    \end{array}\thickspace\right\} \thickspace\thickspace \bm y = {\tt SSM}(\bm u)
\end{equation}
It simply transforms an arbitrary input signal $\bm u$ into an output signal $\bm y$ through some hidden process $\bm x$. Since this model is linear, it can also be phrased as a convolution \cite{lssl, para-lmu} allowing it to be trained in a parallelizable manner without recurrences. More importantly, this model can be made highly efficient and effective by restricting the parameters $\bm A, \bm B, \bm C, \bm D$ to be block-diagonal and ensuring that the transition matrix $A$ is stable i.e., the SSM generates bounded outputs for bounded inputs \cite{s4, s5}. Furthermore, the effectiveness of this model is highly dependent on its initialization. Work has found that this system can encode the history of an input signal effectively with a proper initialization scheme \cite{hippo, trainhippo}. The combination of these ideas culminates in a model called S4 \cite{s4}.

The S4 model is inherently unidirectional. For non-causal applications such as the audio-encoder for offline speech recognition, a bidirectional S4 can be used with non-linear activations and pointwise linear layers \cite{sashimi}:
\begin{equation}
    \label{eq:bissm}
    \begin{alignedat}{3}
        \bm y &\leftarrow {\tt Cat}([{\tt S4}(\bm u), {\tt Rev}({\tt S4}({\tt Rev}(\bm u)))]) \\
        \bm y &\leftarrow {\tt Linear}({\tt Activation}(\bm y))
    \end{alignedat}
\end{equation}
The next section will build upon this model by using multiple parallel heads and introducing an inter-head gating mechanism.

\section{Multi-Head State Space Model}
\label{sec:mssm}

This section will describe a number of technical and architectural proposals for the audio-encoder in the transducer. Section \ref{ssec:mhssm} introduces the stacked MH-SSM. Section \ref{ssec:ihg} describes a novel gating mechanism which allows different SSM heads to communicate. Section \ref{ssec:stateformer} combines the MH-SSM with self-attention for a new transformer architecture. Finally, Section \ref{ssec:msfe} describes how the MH-SSM can be used to replace the convolutional frontend.

\subsection{Stacked \& Multi-Head Extension}
\label{ssec:mhssm}

We extend the S4 layer with a significantly more flexible approach. Taking inspiration from multi-head self-attention, we project the input $D_i$-dimensional signal into $H \in \{2, 4, 8, \dots\}$ separate signals of dimension $\bar{D_i}=D_i/H$, and process each using an independent SSM that is randomly initialized.
\begin{figure}[h]
    \centering
    \includegraphics[width=0.20\textwidth]{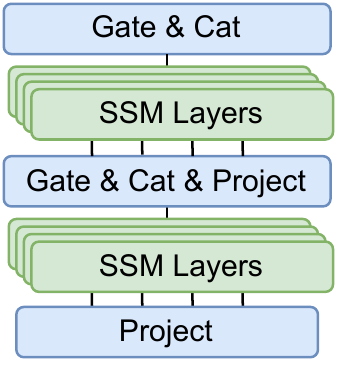}
    \caption{The MH-SSM block first projects an input signal to several lower-dimensional signals, each fed to a separate SSM. The output is gated, concatenated and repeated a second time.}
    \label{fig:mhssm}
\end{figure}
While this can be followed by a simple non-linear activation such as ReLU or GELU, we opt for a novel gating mechanism described in the following section. Finally, we repeat this procedure once again for a stacked module, see Figure \ref{fig:mhssm}. This module would then operate as a drop-in replacement for the $S4$ in Equation \ref{eq:bissm} to form a bidirectional model.
This multi-head design provides the flexibility to learn both meaningful time-steps and different types of temporal dynamics on sequence data.

\subsection{Inter-Head Gating}
\label{ssec:ihg}

\begin{figure}[h]
    \centering
    \includegraphics[width=0.25\textwidth]{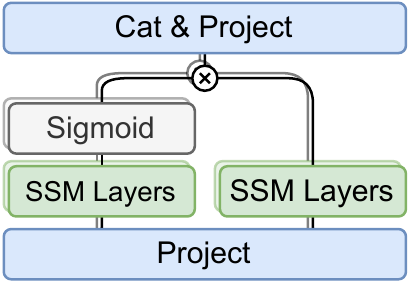}
    \caption{Previous approaches would either apply GELU or GLU to the output of a single SSM. IHG gates the output of an SSM using another independent SSM.}
    \label{fig:ihg}
\end{figure}

By default, prior work has used the GELU activation function, see Equation \ref{eq:bissm}. In some experiments~\cite{s4, sashimi}, the GLU activation was also found beneficial. However, a multi-head state space architecture with $H$ heads offers many more gating possibilities. 
We propose an \textit{inter-head gating} (IHG) approach in which half the number of heads are used to gate the remaining heads, allowing different heads to communicate and generally leading to improved results, as illustrated in Figure \ref{fig:ihg}.
The IHG output is computed by mixing the heads according to (where $\sigma$ refers to the sigmoid):
\begin{equation}
    \label{eq:headgate}
    \begin{alignedat}{3}
        a^{(h)} &= y^{(h)} \cdot \sigma\left(y^{(h + H/2)}\right), \thickspace\thickspace h = \{1, ..., H/2\}
    \end{alignedat}
\end{equation}
allowing different heads to communicate and gate each other, generally leading to improved results. It should be noted that the number of heads $H$ needs to be even. While our approach is wildly different, the notion of allowing heads to communicate has also been suggested in \cite{talkingheads} regarding attention, where linear layers mix information across heads and have been shown to improve performance.

\subsection{Multi-Scale Frontend}
\label{ssec:msfe}

\begin{figure}[!ht]
    \centering
    \includegraphics[width=0.460\textwidth]{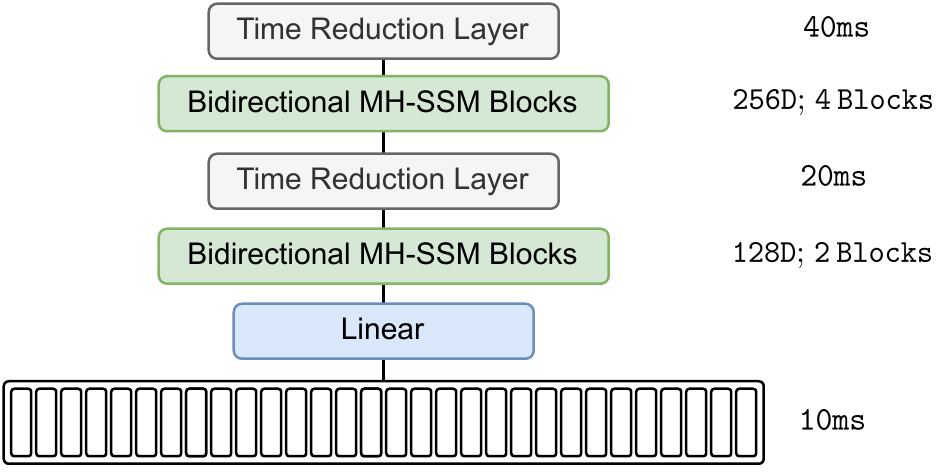}
    \caption{Filterbanks are passed on to a linear layer with 128D outputs. These are fed through 2 smaller MH-SSM residual blocks followed by a time reduction (TR) layer which splices every two frames. This is again followed by another set of MH-SSM blocks and a TR layer resulting in 512D features with an effective 40ms frame rate.}
    \vspace{-10px}
    \label{fig:msfe}
\end{figure}

Audio-encoders typically subsample the input sequence using a convolutional frontend to reduce sequence length and increase computational tractability~\cite{las,rnnt}. In this work, we utilize a multi-scale (MS) state space front end using the MH-SSM to make use of its ability to model longer-range dependencies, see Figure \ref{fig:msfe}. The frontend intertwines MH-SSM blocks that capture temporal dependencies and time reduction layers which reduce the frame rate resulting in the same output size and striding as typical convolutional frontends.

\begin{figure*}[t]
    \centering
    \includegraphics[width=0.95\textwidth]
    {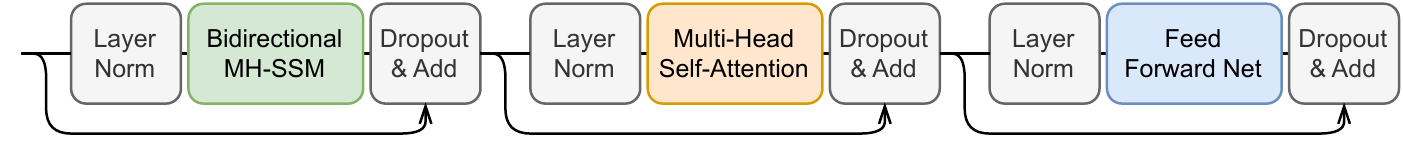}
    \caption{Stateformer: Bidirectional state space augmented transformer block. It simply has an additional block prior to the attention unit with a bidirectional MH-SSM. The pure MH-SSM architecture is similar but without the self-attention block.}
    \vspace{-10px}
    \label{fig:stateformer}
\end{figure*}

\subsection{Stateformer: State Space Augmented Transformer}
\label{ssec:stateformer}

A pure multi-head state space model architecture is attractive due to its ability to capture both short and long-range dependencies. However, since the state space model is equivalent to a linear RNN, it is expressively more limited in the temporal dimension. Therefore, we propose a model combining the MH-SSM with attention by simply inserting a pre-norm bidirectional block prior to the self-attention unit in the transformer architecture, referred to as the \textit{Stateformer}, see Figure \ref{fig:stateformer}. 
%

\section{Experimental Evaluation}

\subsection{Data}
 We evaluate the proposed models on the LibriSpeech dataset \cite{librispeech} consisting of about 960 hours of speech data sampled at 16kHz. 
 SpecAugment~\cite{specaugment} and speed perturbation~\cite{ko2015audio} were used for data augmentation.
 We used a sliding window of 25ms with a 10ms frame shift to extract 80-dimensional filterbank.

\subsection{Baselines}
All speech recognition models were based on the transducer framework \cite{rnnt}, which has three components: encoder, predictor and joiner. We train various transducers by keeping the predictor and joiner fixed, and compare various Transformer \cite{transformer-rnnt}, Conformer \cite{conformer} and S4 \cite{s4} encoders against the proposed approaches. 
The 80-dim input feature was first linearly projected to a dimension of 128.
Furthermore, we explore the style of subsampling frontends by comparing:
\begin{enumerate}[label=(\alph*), leftmargin=5.3mm]
    \item \textit{Time Reduction Layers} (TR): Splicing 128-dim frames to 512 dimensions, reducing the sequence length by 4x.
    \item \textit{Convolutional Layers} (CNN): 2D convolutional network with a total stride of 4 and 512 output channels~\cite{wang2020transformer, conformer}.
    
    \item \textit{Multi-scale Layers} (MS): Proposed multi-scale frontend which intertwines MH-SSM blocks with TR layers, reducing the sequence length by 4x.
\end{enumerate}

\subsection{Implementation Details}
Baseline and proposed models are implemented using an extension of the Fairseq framework~\cite{ott2019fairseq}. The encoder model dimension was set to 512 and kept fixed for all experiments; the model size was controlled by the number of encoder layers. The baselines used the convolutional frontend and consists of 20-36 S4, Transformer or Conformer blocks with 8 self-attention heads (not applicable to S4), and a feed-forward net dimension of 2048. Different to \cite{conformer}, our Conformer baseline did not use a macaron style block as it was not found useful and had the convolutional module prior to the attention with a kernel size of 31. The prediction
network consisted of a three-layer 512-dimensional LSTM with layernorm and dropout. Both the encoder and predictor outputs were projected to 1024 dimensions before being fed into an additive joiner with a single linear layer of $\vert{\mathcal{Y}}\vert = 4097$ sentence-piece~\cite{kudo2018sentencepiece} output units. Similar to the S4 baseline, our proposed MH-SSM simply replaced the self-attention layer of the transformer; the configuration of this layer, such as the number of stacks, heads, use of inter-head gating and frontend subsampler are investigated. The Stateformer uses the same setup as the transformer combined with the best MH-SSM configuration. Large (greater than 100M parameters) baselines and proposed models were also trained using auxiliary classifiers similar to \cite{szegedy2015going}, in which intermediate encoder outputs are trained to predict frame labels every 4 layers.

All models used the Adam optimizer \cite{kingma2014method} with a learning rate linearly warming up to the peak value in 10k iterations, fixed until the 60th epoch and thereafter, exponentially decayed by a factor of 0.96 each epoch. A dropout value of 0.10 is used in all encoders and 0.30 in all predictors and the batch size is set based on occupying maximal GPU memory. All models were trained up to 200 epochs using 32 NVIDIA A100 GPUs. Hyperparameters and level of checkpoint averaging were based on the development set.

\section{Results \& Discussion}

\subsection{Main Results}

Table \ref{tab:main} shows the word error rate performance of our models of the large configuration on LibriSpeech with our baselines and state-of-the-art models including ContextNet \cite{han2020contextnet}, Transformer \cite{transformer-rnnt}, Conformer \cite{conformer, zhang2020pushing} and the recently introduced Branchformer \cite{peng2022branchformer} and E-Branchformer \cite{ebranchformer}. No external language model was used.

\begin{table}[h]
    \centering
    \caption{WER\% performance of baseline and proposed models on Librispeech compared with best results found in the literature (no external language model). At approximately 140.3M parameters, our attention-free MH-SSM model is competitive with ContextNet and outperforms many other reported models. At 139.8M parameters, our Stateformer is competitive with the best-reported Conformer and outperforms all other models.}
    \label{tab:main}
    \resizebox{0.47\textwidth}{!}{
        \begin{tabular}{l|c|cccc}
            \toprule
            \multirow{2}{*}{Model} & \multirow{2}{*}{Params} &  \multicolumn{2}{c}{\tt dev} & \multicolumn{2}{c}{\tt test} \\
            & & clean & other & clean & other \\
            \midrule
            \midrule
            \textbf{\small AED} \\
            Branchformer \cite{peng2022branchformer} & 116.2M & 2.2 & 5.5 & 2.4 & 5.5 \\
            E-Branchformer \cite{ebranchformer} & 148.9M & -- & -- & 2.14 & 4.55 \\
            Conformer \cite {ebranchformer} & 147.8M & -- & -- & 2.16 & 4.74 \\
            \midrule
            \textbf{\small Transducer} \\
            Transformer \cite{transformer-rnnt} & 139M & -- & -- & 2.4 & 5.6  \\
            Transformer \cite{liu2021improving} & 160M & -- & -- & 2.2 & 4.7  \\
            ContextNet~\cite{han2020contextnet} & 112.7M & 2.0 & 4.6 & \bf 2.1 & 4.6 \\
            Conformer \cite{conformer} & 118.8M & \bf 1.9 & \bf 4.4 & \bf 2.1 & \bf 4.3 \\
            Conformer \cite{zhang2020pushing} & $\simeq$120M & 2.0 & 4.7 & 2.2 & 4.8 \\
            \midrule
            \midrule
            \emph{\small Baselines}  & & & & & \\
            S4 36L & 129.6M & 2.21 & 5.63 & 2.41 & 5.68 \\
            Transformer 36L & 129.0M & 2.16 & 5.28 & 2.32 & 5.34 \\
            Conformer 24L & 133.7M & 1.95 & 4.84  & 2.21 & 5.04 \\
            \midrule
            \emph{\small Proposed Models} & & & & & \\
            MH-SSM 32L & 140.3M & 1.80 & 4.96 & 2.01 & 4.61 \\
            Stateformer 25L & 139.8M & \bf 1.76 & \bf 4.37 & \bf 1.91 & \bf 4.36 \\
            \bottomrule
        \end{tabular}
    }
\end{table}

The performance of our (attention-free) multi-head state space model (MH-SSM) is able to achieve competitive results of $1.80/4.96/2.01/4.61$ outperforming existing transformer transducers and competing with ContextNet. It also significantly outperforms the original S4 which is unable to outperform transformer baselines. Furthermore, this model is able to outperform one of the Conformer transducers \cite{zhang2020pushing} on all but dev-other. The Stateformer further pushes the performance by combining MH-SSM blocks with self-attention. The achieved WERs of $1.76/4.37/1.91/4.36$ are highly competitive outperforming essentially all models with one exception, the original Conformer. In this case, the Stateformer is able to outperform the Conformer in all but dev and test-other for which it is competitive. Overall, this demonstrates the power of multi-head state space models both as a standalone model and when combined with self-attention blocks.

\subsection{Ablation Studies}

\subsubsection{Baselines}

Table \ref{tab:baseline} reports the WER performance of smaller baseline models. Overall we can observe that the attention-free S4 (SSM baseline) can outperform the transformer with a time reduction frontend. With a convolutional frontend the larger Conformer is best followed by the transformer. The transformer benefits more from a convolutional frontend as it complements self-attention. Models with convolutional aspects, Conformer and S4 (which can be seen as a variable length convolution) benefit less.

\begin{table}[h!]
    \centering
    \caption{Baseline S4, Transformer and Conformer WER\% performance with various frontends (FE). With a CNN frontend, a larger Conformer is best followed by the Transformer.}
    \label{tab:baseline}
    \resizebox{0.47\textwidth}{!}{
        \begin{tabular}{l | cc | cccc}
            \toprule
            \multirow{2}{*}{Model} & \multirow{2}{*}{Params}  & \multirow{2}{*}{FE} & \multicolumn{2}{c}{\tt dev} & \multicolumn{2}{c}{\tt test} \\
            & & & clean & other & clean & other \\
            \midrule
            \midrule
            \multirow{2}{*}{S4 20L} & 78.1M & TR & 2.45 & 6.88 & 2.71 & 6.72 \\
            & 78.8M & CNN & 2.36 & 6.60 & 2.67 & 6.47 \\
            \midrule
            \multirow{2}{*}{Transformer 20L} & 76.7M & TR & 2.96 & 7.09 & 3.05 & 7.18 \\
            & 77.5M & CNN & 2.45 & 5.82 & 2.62 & 6.15   \\
            \midrule
            \multirow{2}{*}{Conformer 20L} & 91.9M & TR & 2.17 & 5.54 & 2.43 & 5.45 \\
            & 92.7M & CNN & 2.04 & 5.31 & 2.26 & 5.37 \\
            \bottomrule
        \end{tabular}
    }
\end{table}
\vspace{-10px}
\subsubsection{Stacking SSM Layers}


There are a number of differences between a standard S4 and the MH-SSH block, specifically, the stacking of SSM layers within a single residual block. Table \ref{tab:stack} shows the contrast between stacking SSM layers within a residual block versus opting for deeper models. Stacking was found marginally better, and more importantly, allows the model to scale to larger sizes.

\begin{table}[h!]
    \centering
    \caption{Impact of stacking state space layers in a residual block. Stacking 2 layers was found to be marginally more effective than increasing the number of layers.}
    \label{tab:stack}
    \resizebox{0.47\textwidth}{!}{
        \begin{tabular}{c | cc | cccc}
            \toprule
            \multirow{2}{*}{Layers} & \multirow{2}{*}{Params}  & \multirow{2}{*}{Stack} & \multicolumn{2}{c}{\tt dev} & \multicolumn{2}{c}{\tt test} \\
            & & & clean & other & clean & other \\
            \midrule
            \midrule
            \multirow{2}{*}{16} & 56.8M & 1 & 2.57 & 7.13 & 2.79 & 6.86  \\
            & 66.3M & 2  & \bf 2.36 & \bf  6.88 & \bf 2.52 & 6.59  \\
            \midrule
            \multirow{1}{*}{20} & \multirow{1}{*}{67.6M}
            &  1 & 2.42 & 6.92 & 2.67 & \bf 6.56\\
            \bottomrule
        \end{tabular}
    }
\end{table}

\subsubsection{Number of Heads and IHG}

Next, we compare the number of SSM heads and the impact of using IHG instead of standard GLU activations, see Table \ref{tab:heads}. All of the MH-SSM models use a simpler TR frontend. At 74.7M parameters, the 4H with IHG model is able to outperform the 20L transformer baseline and rival the one with CNN frontend. Overall, the table shows the effectiveness of combining multi-head with head gating.

\begin{table}[h!]
    \centering
    \caption{Ablation study investigating the number of heads and use of IHG in the MH-SSM model. All models use the TR frontend and have 74.7M parameters.}
    \label{tab:heads}
    \resizebox{0.37\textwidth}{!}{
        \begin{tabular}{cc | cccc}
            \toprule
            \multirow{2}{*}{$\#$H}  & \multirow{2}{*}{IHG} & \multicolumn{2}{c}{\tt dev} & \multicolumn{2}{c}{\tt test} \\
            & & clean & other & clean & other \\
            \midrule
            \midrule
            2 & \xmark & 2.33 & 6.92 & 2.58 & 6.49 \\
            4 & \xmark & 2.23 & 6.74  & 2.52 & 6.46 \\
            2 & \cmark & 2.13 & 6.81 & 2.47 & 6.44 \\
            4 & \cmark & 2.19 & 6.38 & 2.42 & 6.25 \\
            8 & \cmark & 2.17 & 6.49 & 2.43 & 6.20 \\
            \bottomrule
        \end{tabular}
    }
\end{table}

\subsubsection{Multi-Scale Frontend}

Using the best found MH-SSM from the previous section with 4 heads and IHG enabled, we evaluate the impact of including a multi-scale frontend, see Table \ref{tab:fes}. With a minor increase of 4.5M parameters, the performance of the MS + MH-SSM system is significantly better. The corresponding Stateformer further improves performance by a notable margin, outperforming the Conformer baseline in Table \ref{tab:baseline}.

\begin{table}[h!]
    \centering
    \caption{Simple comparison of TR vs MS frontend (FE) for the MH-SSM model; including a final Stateformer based on the best found MH-SSM model.}
    \label{tab:fes}
    \resizebox{0.47\textwidth}{!}{
        \begin{tabular}{l | cc | cccc}
            \toprule
            \multirow{2}{*}{Model} & \multirow{2}{*}{Params}  & \multirow{2}{*}{FE} & \multicolumn{2}{c}{\tt dev} & \multicolumn{2}{c}{\tt test} \\
            & & & clean & other & clean & other \\
            \midrule
            \midrule
            \multirow{2}{*}{MH-SSM 16L} & 74.7M & TR & 2.19 & 6.38 & 2.42 & 6.25 \\
            & 79.2M & MS & 2.14 & 6.12 & 2.39 & 5.99 \\
            \midrule
            Stateformer 16L & 96.1M & MS & 2.06 & 5.01 & 2.27 & 5.07 \\
            \bottomrule
        \end{tabular}
    }
\end{table}

\section{Conclusions}

We proposed a multi-head state space model (MH-SSM) for the audio-encoder of transducer-based speech recognition. Parallel heads of SSMs, together with a novel inter-head gating mechanism was shown highly effective yielding a new class of high-performing models. In addition, we combine MH-SSM blocks and self-attention to form a new type of transformer architecture--the Stateformer. On the LibriSpeech speech recognition task, the proposed MH-SSM outperforms transformer baselines by a large margin. Furthermore, the Stateformer further improves the performance, achieving state-of-the-art performance without external language models.

\clearpage
\bibliographystyle{IEEEtran}
\bibliography{mybib}
\end{document}